\documentclass[twocolumn,prl,superscriptaddress]{revtex4}

\usepackage[latin1]{inputenc}
\usepackage[T1]{fontenc}
\usepackage{lmodern}
\usepackage{eurosym}
\usepackage[frenchb, english]{babel}
\usepackage{amsmath}
\usepackage{amsthm}
\usepackage{amssymb}
\usepackage{varioref}
\usepackage{listings}
\usepackage{graphicx}
\usepackage{float}
\usepackage{color}
\usepackage[usenames,dvipsnames]{xcolor}
\usepackage{epstopdf}
\usepackage{doublestroke/dsfont}

\labelformat{Thm}{Theorem~#1}

\newcommand \be  {\begin{equation}}
\newcommand \beno  {\begin{equation*}}
\newcommand \bea {\begin{eqnarray} \nonumber }
\newcommand \ee  {\end{equation}}
\newcommand \eeno  {\end{equation*}}
\newcommand \eea {\end{eqnarray}}

\let\a=\alpha \let\b=\beta  
   \let\k=\kappa
    
  \let\f=\varphi 
   
 \let\Th=\Theta  

 \let\th=\theta \let\io=\infty

\def\Re{{\rm Re}\,}\def\Im{{\rm Im}\,}

\def\de{\mathrm d}

\begin{document}

\title{Endogenous crisis waves: a stochastic model with synchronized collective 
behavior}

\author{Stanislao Gualdi}
\affiliation{Universit\'e Pierre et Marie Curie - Paris 6, Laboratoire de Physique 
Th\'eorique de la Mati\`ere 
Condens\'ee, 4, Place Jussieu, 
Tour 12, 75252 Paris Cedex 05, France}
\affiliation{IRAMIS, CEA-Saclay, 91191 Gif sur Yvette Cedex, France}
\author{Jean-Philippe Bouchaud}
\affiliation{CFM, 23 rue de l'Universit\'e, 75007 Paris, France, and Ecole 
Polytechnique, 91120 Palaiseau, France}
\author{Giulia Cencetti}
\affiliation{Laboratoire de Physique Th\'eorique,
\'Ecole Normale Sup\'erieure, UMR 8549 CNRS, 24 Rue Lhomond, 75231 Paris Cedex 
05, France}
\author{Marco Tarzia}
\affiliation{Universit\'e Pierre et Marie Curie - Paris 6, Laboratoire de Physique 
Th\'eorique de la Mati\`ere 
Condens\'ee, 4, Place Jussieu, 
Tour 12, 75252 Paris Cedex 05, France}
\author{Francesco Zamponi}
\affiliation{Laboratoire de Physique Th\'eorique,
\'Ecole Normale Sup\'erieure, UMR 8549 CNRS, 24 Rue Lhomond, 75231 Paris Cedex 
05, France}

\begin{abstract} 
We propose a simple framework to understand commonly observed crisis waves in 
macroeconomic Agent Based models, that is also relevant to a variety of other physical or biological situations
where synchronization occurs. We compute exactly the phase diagram of the model and the location of the synchronization transition in parameter
space. Many modifications and extensions can be studied, confirming 
that the synchronization transition is extremely robust against various sources of noise or imperfections. 
\end{abstract}

\sloppy
\maketitle

Synchronisation is arguably among the most baffling cooperative phenomenon 
in nature \cite{Sync}. Examples abound in physics and chemistry, but biological 
realizations are probably the most relevant to us %humans 
as they involve in
particular the synchronization of pacemaker cells in the heart or of neurons 
firing 
in the brain. Synchronization of clapping in concert halls or of flashing in 
assemblies of thousands of fireflies are also well known. The latter case is 
particularly interesting: although the effect is so to say visible to the naked 
eye,
its very existence is so counter-intuitive that for several decades no one could 
come
up with a plausible theory. As vividly recounted by Strogatz \cite{Sync}, 
scientists swayed between denial (assigning
the flashing to a periodic quiver of the eyelids of the observer, or to the 
unavoidable 
presence of a ``maestro'') and explanations that were ``more remarkable than the 
phenomenon
itself'' \cite{Sync}. The idea that interacting oscillators can {\it 
generically} synchronize only 
slowly emerged at the end of the sixties \cite{Winfree}, before becoming a 
well-established mathematical 
truth with the work of Kuramoto \cite{Kuramoto}, Strogatz \& collaborators 
\cite{Strogatz}, and many others (for a review, see
\cite{Ritort}). 

\begin{figure}
\includegraphics[scale=0.3]{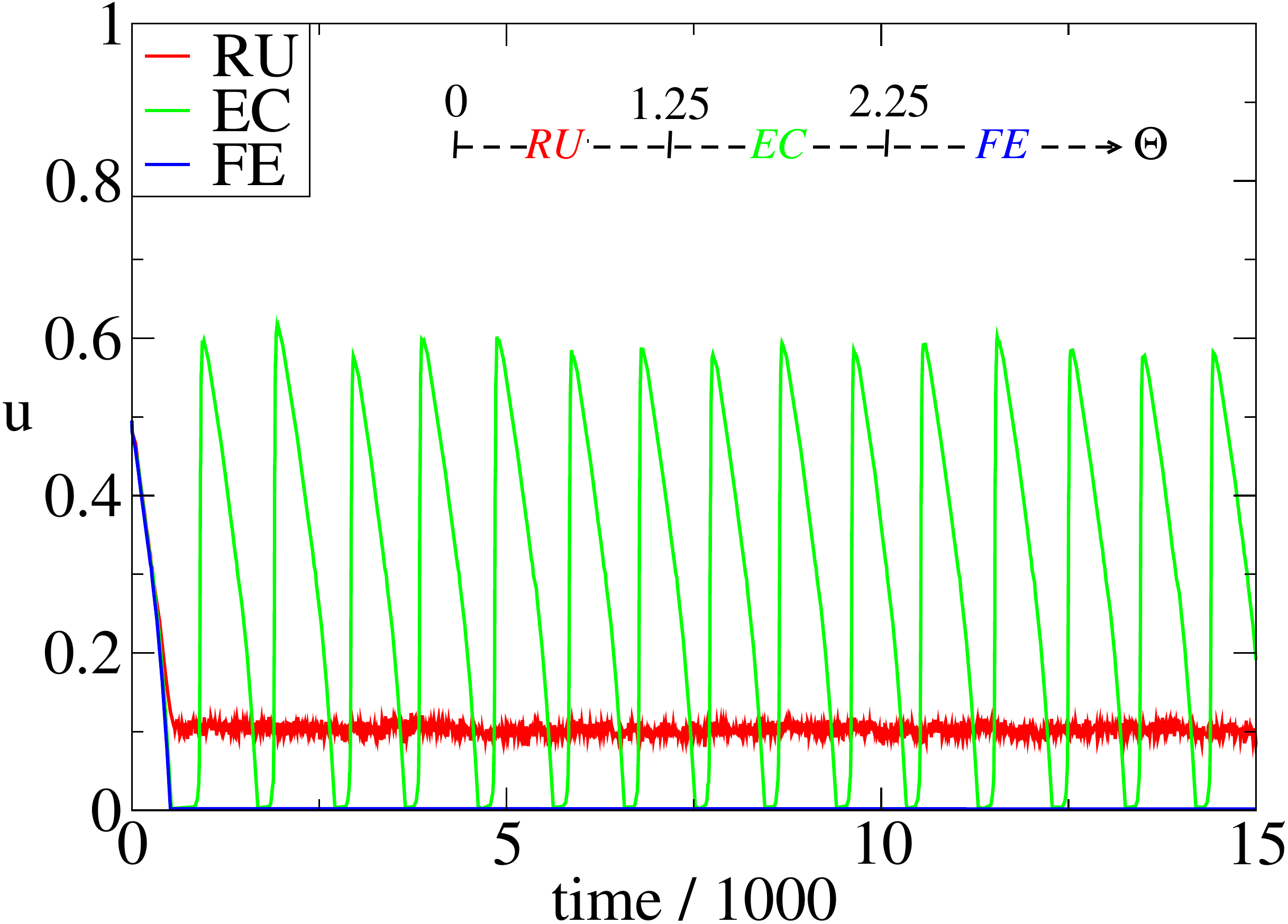}
\caption{Typical trajectories of the unemployment rate $u$ as a function of time 
for the macroeconomic agent based model described in~\cite{Gualdi}, 
for different bankruptcy thresholds $\Theta$. As $\Theta$ is increased, the
economy evolves from a phase of residual unemployment (RU), endogenous crises (EC) and 
finally full employment (FE). In the EC phase, surprisingly, one observes nearly periodic 
bursts of unemployment, corresponding to collective waves of bankruptcies.}
\label{fig:fig1}
\end{figure}

Still, the phenomenon is so unexpected (at least for those only vaguely 
acquainted with these results) that
while working on a stylized agent based model (ABM) of the macroeconomy 
\cite{Gualdi}, we first disbelieved our results. 
We found a whole region of parameter space where the dynamics appeared
to be (see Fig.~\ref{fig:fig1}) a nearly periodic succession of eras of 
prosperity interrupted by acute crisis of
purely endogenous origin -- indeed, no adverse exogenous shocks are present in our model; 
for details, see \cite{Gualdi}. 
In view of the amount of heterogeneity and randomness in our model (firms in our 
economy are all different, 
bankrupted firms are revived at Poisson random times, etc.), such a regular 
succession of crises waves
is both interesting and surprising, and begs for a convincing theoretical 
explanation. 
However, although highly stylized, our ABM is too complex to be amenable to an 
exact analytical treatment, 
so further simplification is needed. This led us to the bare-bones model 
described below, which
indeed exhibits -- in fact, to our surprise -- a synchronization transition. This 
model can be analyzed using reasonably straightforward mathematical tools. 
For instance the location of the transition in parameter space can be computed exactly.  
Many modifications and extensions of the model can be studied, exactly or numerically. Their analysis 
confirms that the synchronization transition is extremely robust, in particular 
against various sources of
noise and imperfections.

Our model is an alternative to the well-known Kuramoto model that has been 
studied inside-out \cite{Kuramoto,Strogatz,Ritort}, and which 
starts from the assumption that isolated individual elements are oscillators. 
Our setting is directly
motivated from the macroeconomic model we wanted to elucidate, and is in fact 
close in spirit to 
``integrate-and-fire'' models of neurons \cite{Strogatz}. It can also be 
rephrased as a mean-field model for epidemic dynamics \cite{Epidemic}, 
fiber bundles~\cite{Hansen}, depinning of elastic manifolds~\cite{Depinning}, or interbank default contagion~\cite{Battiston}. 
In order to describe the model, we choose to
keep with the vocabulary of our macroeconomic model -- transposition to other 
contexts is quite transparent and 
will be discussed below. Each firm $i=1,\dots, N$ is characterized by its 
``financial 
fragility'' measured by the ratio $x_i$ of its outstanding debt to total assets. 
We use the convention that $x_i < 0$ for
indebted firms, and $x_i > 0$ for cash/loan-rich firms. In the course of time, 
this ratio
can increase or decrease due to the success of its business, its needs for cash, 
etc. We posit \cite{Gualdi} that when the
financial fragility exceeds a certain threshold $\Theta$ (i.e. when $x_i \leq - 
\Theta$) banks are reluctant to 
restructure the debt of the firm, which then files for bankruptcy. (In a first 
version of the model, we assume this 
occurs with probability $1$ as soon as $x_i \leq - \Theta$, but one can also consider 
the case where bankruptcy is not 
certain and happens with probability $\kappa \mathds{1}_{x_i \leq - \Theta}$ per 
unit time, see below). Until firms hit
the threshold, the evolution of $x_i$ is modeled as a biased random walk. The 
most important feature of the model 
is the feedback between bankruptcies and the drift of these random walks. In the 
absence of money creation (i.e. 
if the banking sector does not act as a buffer), the outstanding debt of the 
defaulted firm is spread between 
remaining firms -- thereby directly increasing their financial fragility -- and the 
household sector, leading to a decrease 
of its purchasing power and hence worsened business conditions for the surviving 
firms (see~\cite{Gualdi} for a concrete implementation of this general idea). 
In both cases, this gives a negative
contribution to the drift of $x_i$'s. Finally, bankrupted firms are either 
revived or replaced by new firms at a 
certain rate $\varphi$ per unit time, and start with a zero initial fragility.

Taking the limit of a large number of firms, the above rules translate into the 
following Fokker-Planck 
equation for the probability density $P(x,t)$ of observing firms with a certain 
fragility between $x$ and $x+ {\rm d}x$ at time $t$:
\be\label{eq:FP}
\dot{P}(x,t) = DP''(x,t) + b(t) P'(x,t) + J(t)\delta(x-\Theta),
\ee
where we made a translation $x\to x+\Theta$, $\phi(t) = \int_0^{\infty} {\rm d}x 
P(x,t)$ is the fraction of active firms at time $t$, 
$J(t)$ is the flux of new-born firms at time $t$. The boundary condition 
$P(0,t)=0, \forall t$ corresponds to case where bankruptcy is immediate, 
as soon as the threshold is reached, and 
\be\label{eq:FP2}
b(t) = b +  \beta D P'(0,t) \Theta \,
\ee
is the drift, that itself depends on the flux of bankruptcies $D P'(0,t)$ 
(i.e. the number of firms that cross $x=0$ at time $t$ that spread 
an extra debt $\Theta$) through a phenomenological, adimensional parameter 
$\beta$ that measures the strength of the feedback. Note that with our sign convention,
a positive $b(t)$ corresponds to a drift towards {\it more negative} values of $x$.
The reinjection current $J(t)$ is either (model I) a time-independent constant 
$J(t) = \f$, modeling the fact that new firms are created at a constant 
rate, independently of the number of existing firms, or (model II) given by 
$J(t) = (1-\phi(t))\varphi$, meaning that bankrupted firms have a probability $\varphi$ 
per
unit time to be revived, in line with the choice made in the macroeconomic 
model studied in \cite{Gualdi}.  

In both cases
the model contains 5 parameters: $D,b,\varphi,\Th$ and the feedback parameter 
$\beta$, but two can be fixed by the choice of units of time $t$ and ``length'' $x$. 
This leaves us with three adimensional parameters: $\beta$, the Peclet number 
$Pe = b \Th/D$ that compares drift to diffusion and $z = b/(\Th \varphi)$ that 
compares
the average revival time to the time needed to travel $\Theta$ under the action 
of the drift $b$. $Pe \gg 1$ means that diffusion is a small correction to 
drift in typical trajectories, whereas $z \ll 1$ means that revival is fast 
compared to the time to bankruptcy due to drift only.   
As implicitly assumed in the above discussion, we will only consider the case $b 
 > 0$ (i.e. a negative drift on $x$), 
that allows the existence of a stationary state in the absence of feedback 
($\beta=0$). This also describes the economy considered in \cite{Gualdi}: 
for non storable goods and in the absence of productivity gains or innovation, 
our 
myopic firms have a tendency to over-produce and lose money on average. 

We therefore look for a stationary state $P_0(x)$, characterized 
by a constant bankruptcy flux $J(t) = J_0$ and constant $\phi(t) = \phi_0$. 
Because $D P_0'(x=0)$ is the out-coming flux, and the fraction of active firms $\phi$ is time independent, 
it follows that $D P_0'(x=0) = J_0$ and this leads to a constant drift $b_0 = b + \beta \Theta 
J_0$. By setting $\dot{P} \equiv 0$ in Eq. (\ref{eq:FP}), it is 
easy to show that:
\be\label{Puno}
P_0(x) = \frac{J_0}{b_0} \times \begin{cases}
(1 - e^{-\hat x}) & \text{for } \hat x < \hat \Th \ , \\
(e^{\hat \Th} -1 ) e^{-\hat x} & \text{for } \hat x > \hat \Th \ , \\
\end{cases}
\ee
where $\hat{x}= b_0x/D$ and $\hat{\Theta}= b_0\Theta/D$ are adimensional 
``lengths''. The stationary solution for $P_0(x)$ is therefore 
given in terms of the unknown parameters $J_0, \phi_0$ and $b_0$ that are 
determined from:
%\footnotesize
\be
\label{Ptre}
\phi_0  \equiv \int_0^\io dx P_0(x) = \frac{J_0 \Th}{b_0}; \qquad b_0  = b + \beta 
J_0 \Theta.
\ee
%\normalsize
For model I, one has $J_0 = \f$ and $\phi_0$ and $b_0$ are immediately obtained.
For model II, $J_0  = (1 - \phi_0) \varphi$ gives a third equation. 
Using Eqs. (\ref{Ptre}),
we can solve for $b_0 = \frac{ b }{1-\b \phi_0}$ and 
$J_0 = \frac{\phi_0}{\Th} \frac{ b }{1-\b \phi_0}$ and obtain a self-consistent 
equation
\be
\label{eq:phi0}
\frac{\phi_0}{\Th} \frac{b}{1 - \beta \phi_0} = (1 - \phi_0) \varphi  \ , 
\ee
which leads to a second degree equation for $\phi_0$. If the time needed for 
revival is much shorter than the lifetime of the firms, $z \to 0$, 
we expect that $\phi_0 \to 1$, at least for small $\beta$'s. This allows one to 
choose the  correct sign of the solution:
\small
\be
\label{sol:phi0}
\phi_0 =\frac{1}{2 \beta} \left[ 1 + \beta + z - \sqrt{ z^2 + 2z (1+\beta) + (1 
- \beta)^2 } \right] \ .
\ee
\normalsize
It can easily be checked that this solution is then {\it always} such that 
$\phi_0 \in [0,\min(1,1/\beta)]$. Therefore, the stationary solution 
always exists, and $b_0$ never diverges, which would be an obvious sign of an 
instability (see below for a case where this divergence actually happens). 
Still, the question is whether this stationary solution is dynamically stable. 
We therefore write:
\be
P(x,t)= P_0(x) + \epsilon P_1(x,t),
\ee
which induces corresponding $O(\epsilon)$ corrections to $\phi(t)=\phi_0 + \epsilon \phi_1(t)$ and similarly 
for $J(t)$ and $b(t)$. To order $\epsilon$, the diffusion equation for $P$ becomes, both for models I and II:
%\small
\bea
\label{eqP1}
[G^{-1} P_1](x,t) &\equiv& \dot{P}_1(x,t) - DP_1''(x,t) - b_0 P_1'(x,t) \\
&=& b_1(t) P_0'(x) - \f \phi_1(t) \delta(x-\Theta) \,
\eea
%\normalsize
where $G$ is the propagator of random walks with a wall at $x=0$ and drift 
$b_0$, given by the method of images \cite{Redner}:

 \bea
 G(x,t|y,t-\tau) &=& \frac{1}{\sqrt{4 \pi D\tau}} 
 \left\{ \exp\left[-\frac{(x-y+b_0\tau)^2}{4D\tau}\right]\right. \\
  &-& \left. e^{b_0y/D} \exp\left[-\frac{(x+y+b_0\tau)^2}{4D\tau}\right]\right\}
 \eea

\begin{figure}
\includegraphics[scale=0.3]{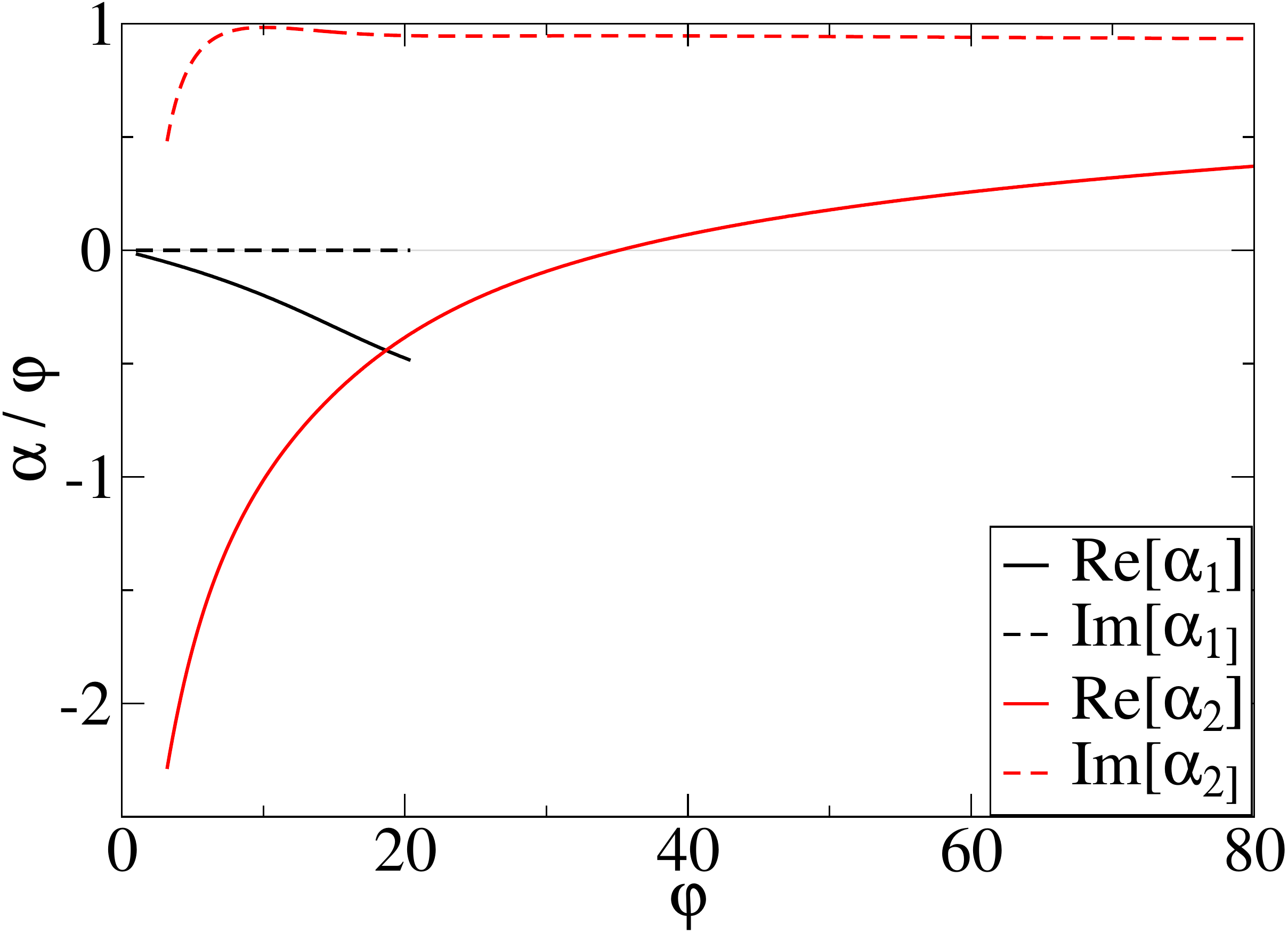}
\includegraphics[scale=0.3]{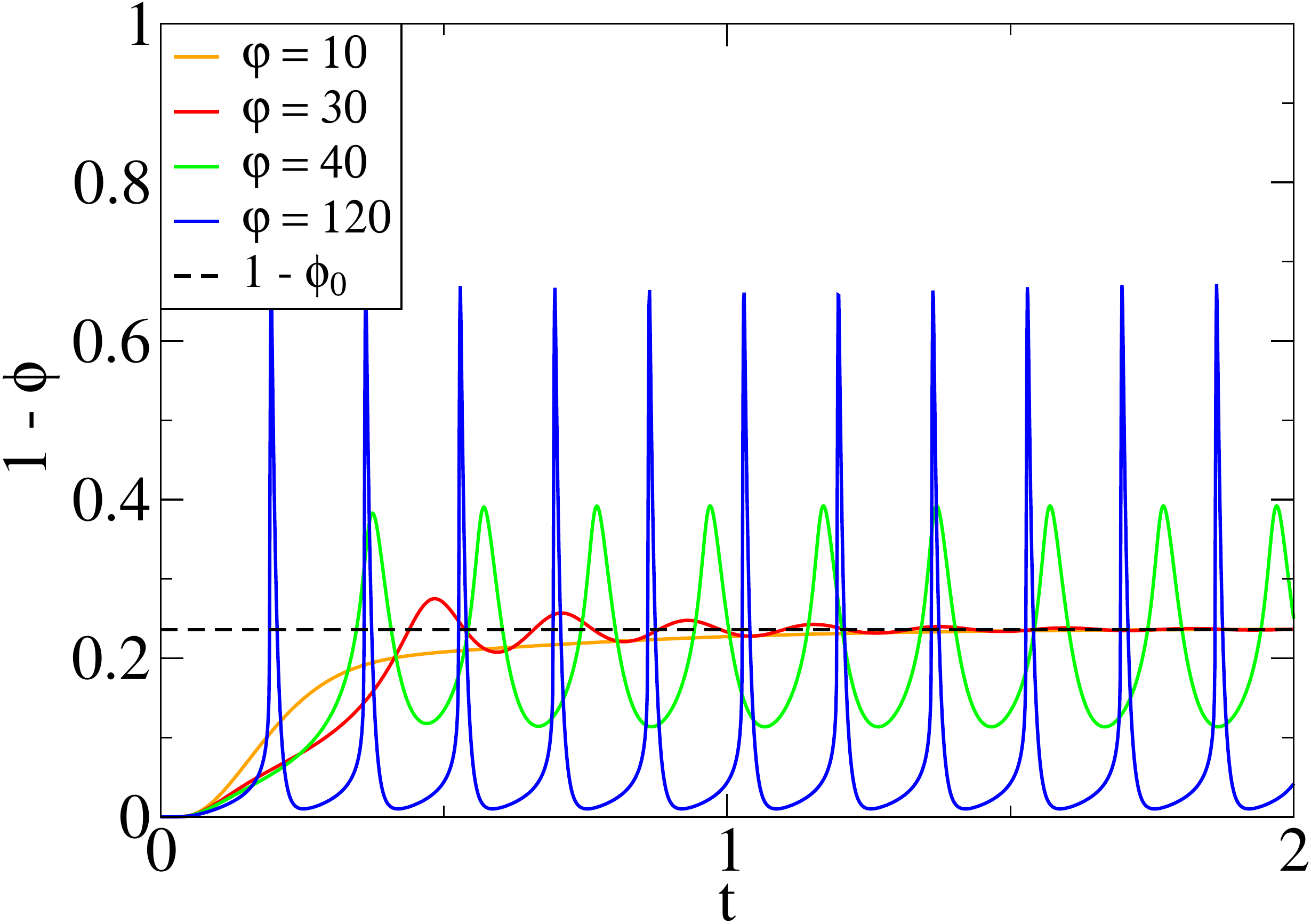}
\caption{(\emph{Top}) Numerical solutions of Eq.~\eqref{eq:alpha}: for small 
values of $\f$ there is only one linearly stable solution 
without oscillations ($\alpha_1$); for intermediate values of $\f$ a second linearly stable 
solution appears with oscillations ($\alpha_2$); for high values of 
$\f$ there is only a linearly unstable solution with oscillations 
corresponding to the appearance of the synchronized behavior.
(\emph{Bottom}) Typical trajectories
of the fraction of bankrupt firms $1-\phi$ as a function of time for different values
of $\varphi$. $\phi_0$ is the theoretical value of stationary fraction of active firms, given
by Eq. (\protect{\ref{sol:phi0}}). For $\varphi>\varphi_c$ the synchronized behavior
settles in. The plot corresponds to $z=0.002$ and $\beta=1.3$. 
}
\label{fig:fig2}
\end{figure}

\begin{figure}
\includegraphics[scale=0.3]{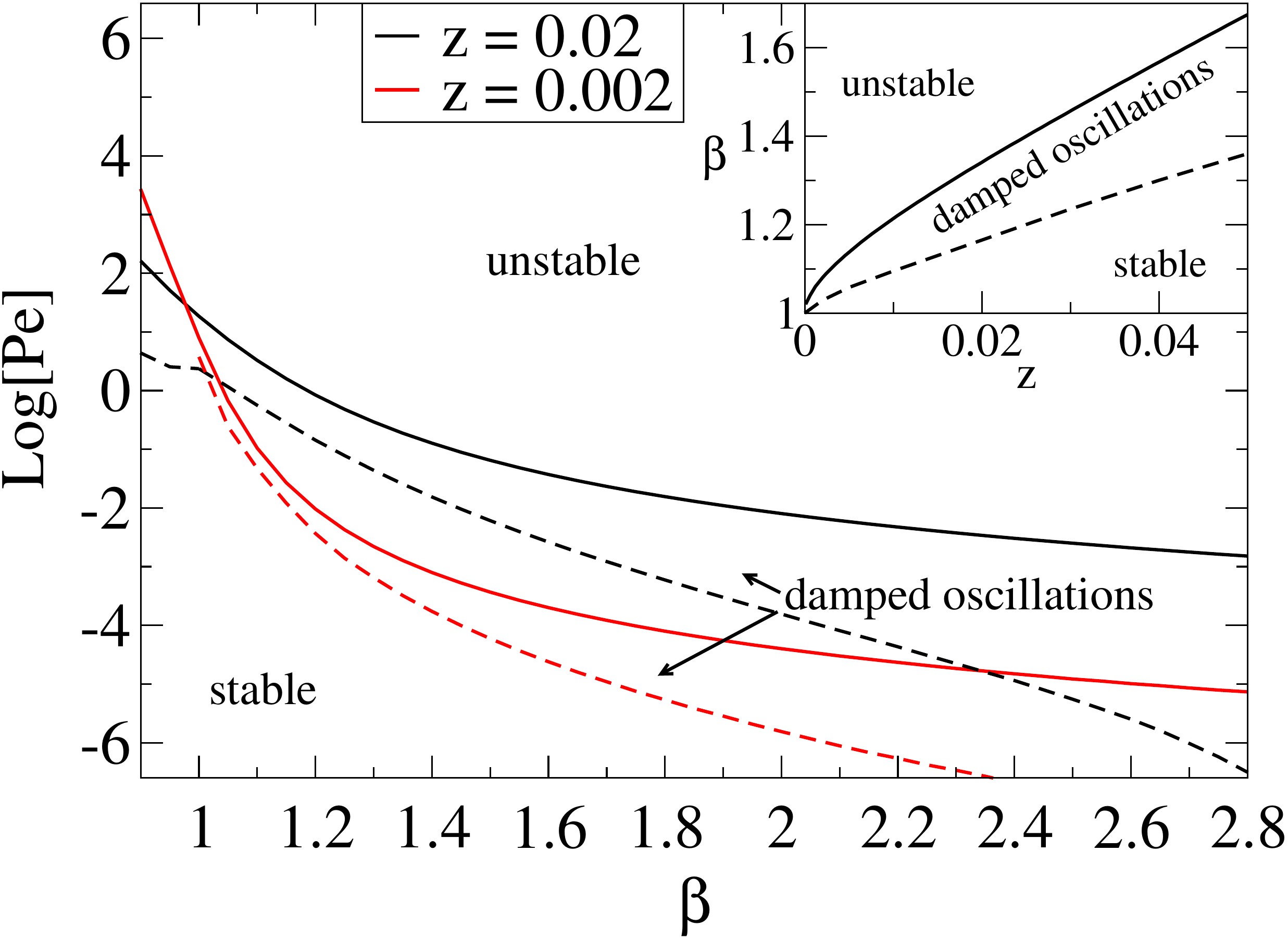}
\caption{Phase diagram of the model in Eq.~\eqref{eq:FP} in the $(\beta,Pe)$ plane
as given by the solutions of Eq.~\eqref{eq:alpha} with $z=0.02$ (black lines) and $z=0.002$ (red lines).
Dashed lines separate the region where the maximal solution has $\Re(\a)<0$ and $\Im(\a)=0$ (stable)
from the region where $\Re(\a)<0$ and $\Im(\a)\neq0$ (damped oscillations). Full lines separate the latter region
from the one where $\Re(\a)>0$ (unstable). In the inset we plot the phase diagram in the $(\beta,z)$ for $Pe=0.5$.
}
\label{fig:fig3}
\end{figure}

Therefore, formally:
\be
\label{eqP1bis}
\begin{split}
P_1(x,t) &= \int_0^\infty d\tau \int_0^\infty dy G(x,t|y,t-\tau)\\
&\times \left[b_1(t-\tau) P_0'(y) - \f \phi_1(t-\tau) \delta(y-\Theta)\right].
\end{split}
\ee
This expression gives $P_1(x,t)$ in terms of $b_1(t)$ and $\phi_1(t)$. Using the 
definition of $b(t)$ and $\phi(t)$ then
leads to self-consistent dynamical equations for $b_1(t)$ and $\phi_1(t)$, which
are slightly different in models I and II. 
We focus on model II and
we assume (and check self-consistently) that $\phi_1(t) = \Phi e^{\alpha t}$ 
and $b_1(t) = B \Theta e^{\a t}$, where $\alpha$ is (a priori) a complex number 
with $b_0^2 + 4 D \Re(\a) \geq 0$ to have convergent integrals.
After simple algebra, one finally finds that $B$ and $\Phi$ have to obey the 
following equations: 
\bea\label{eqexp:uno}
\Phi &=& - \frac{B \phi_0}{\a}  \frac{1 -e^{K\Th}}{1 -K D/b_0} 
- \f \Phi  \frac{1 - e^{ K \Th} }{\a};\\
B &=& -  \b \f \Phi e^{K \Th} +     B \b \phi_0 \frac{1 -e^{K\Th}}{1 -K D/b_0} 
\ ,
\eea
where $K = \frac{b_0}{2 D} \left(1 - \sqrt{1 + \frac{4\a D}{b_0^2} } \right)$.
These have non-zero solution only if $\a$ satisfies the equation \cite{footnote}:
\be\label{eq:alpha}
\b\phi_0 (\a + \f) = [ \a + \f (1- e^{K \Th}) ]  \frac{1 - K D/b_0}{1-e^{K \Th}} 
\ .
\ee
This equation has to be solved numerically in the complex plane, 
and one has to choose the solution with
the largest $\Re(\a)$ that dominates the long-time evolution.
We find three different 
possibilities: $\Re(\a) <  0, \Im(\a)=0$: the stationary state $P_0$ is linearly
stable and relaxation towards it is exponential; $\Re(\a) <  0, \Im(\a) \neq 0$: 
the stationary state $P_0$ is linearly
stable and relaxation towards it is exponential with oscillations; $\Re(\a) >  
0, \Im(\a) \neq 0$: $P_0$ is linearly unstable and the oscillations are the
precursor of the synchronized state observed numerically. An example of the 
dependence of $\a$ as a function of $\varphi$ for fixed $\beta$ and  $z$ is given in 
Fig.~\ref{fig:fig2}-a.
One sees that $\Re(\a)$ crosses zero continuously for a critical value 
$\varphi_c$, for which $\Im(\a) \neq 0$. The corresponding evolution of fraction 
of 
{\it inactive} firms, $1 - \phi(t)$, is given in 
Fig.~\ref{fig:fig2}-b, where we show a numerical 
integration of the Fokker-Planck equation, Eq. (\ref{eq:FP}), but identical results
are obtained from a random walk simulation of $10^5$ firms obeying the same dynamics.
One clearly 
sees that the synchronized behavior sets in exactly at the value 
$\varphi=\varphi_c \approx
35.15$ (for $\beta=1.3$ and $z=0.002$) predicted by the linear stability 
analysis. The transition is found to be continuous, with an amplitude of the oscillations
that vanishes at the transition, and a frequency given by $\Im(\a_c)$. The phase diagrams in the $(Pe, \beta)$ plane (for a fixed $z$) and 
in the $(z, \beta)$ plane (for a fixed $Pe$) are given in Fig.~\ref{fig:fig3}. The conclusion of 
our analysis is that crises waves and synchronization indeed appear in our skeleton macroeconomic 
model\footnote{In order to understand the ``reentrant'' nature of the phase diagram shown in Fig. 1 -- i.e. the fact that the system is 
stable, then unstable and stable again as $\Theta$ is increased, one should bear in mind that 
the values of $\beta,Pe,z$ needed to describe the macroeconomic ABM of \cite{Gualdi} all depend on $\Theta$ itself. 
One expects that in the prosperous FE phase, the feedback parameter $\beta$ is reduced, which indeed leads to a disappearance of the
oscillations.}. What is quite non trivial is that this transition survives the fact that 
firms perform {\it independent} Brownian motion and are reinjected in the system 
at random Poisson time. Similar conclusions of course hold 
for the Kuramoto model as well.  

The above stability analysis can be extended in different directions. First, the 
reinjection flux can be taken as a constant $J_0 = \f$ (Model I), leading to an even more
unstable system (see \cite{footnote}). Second, the reinjection flux does not need to
be localized on $x=\Th$ but it can be spread out over a certain region, i.e. one can 
replace the
term $J(t)\delta(x-\Theta)$ in Eq. (\ref{eq:FP}) by $J(t)f(x-\Theta)$, where $f(x)$ is a function peaked in zero, normalized to one,
and with finite width $w$,
with only 
quantitative changes. In fact, starting with a $\delta$ function in the 
synchronized phase, one can induce the transition by increasing the width $w$ 
beyond some critical value $w_c$.
Third, one can replace the deterministic
bankruptcy condition by a stochastic one, by adding a term $-\kappa \th(-x)$ to 
the Fokker-Planck equation and setting 
$b(t) = b +  \beta \k \int_{-\infty}^0 \de x (\Th - x) P(x,t)$. The above case 
corresponds to the limit $\kappa = \infty$. One finds that the synchronization
phenomenon survives at finite $\kappa$. For example, when $\beta=1.3, z=0.002, 
\varphi=80$, synchronization occurs for $\kappa > \kappa_c \approx 100$. It is
important (but again not very intuitive) that the synchronization phenomenon 
does not sensitively depend on the presence of a well defined threshold -- one 
does
not expect neurons or fireflies to be perfectly tuned to a precise firing 
threshold. Finally, we have investigated the case where the bankruptcy feedback 
depends 
on the number of {\it active} firms, i.e. $b(t) = b +  \frac{\beta}{\phi(t)} D 
P'(0,t) \Theta$, corresponding to the case where the debt of the failing firms 
is 
spread among the surviving firms only. In this case, we find that the above 
stationary state $P_0(x)$ ceases to exist as soon as $\beta > 1$, which becomes 
the
threshold for synchronized behavior. In this case, the transition is found to 
be first order.

Finally, we want to mention different potentially interesting interpretations of 
our model. First, interbank default contagion, which has become a major theme 
since the 2008 crisis \cite{Battiston}. Here, the translation is almost 
immediate: since the assets of one bank is the liability of another, the default 
of one bank reduces the
equity of its lenders, therefore pushing themselves closer to default. Treating 
the model in mean-field immediately leads to a Fokker-Planck equation like 
Eq.(\ref{eq:FP}). In a recent study of a similar model, J. Bonart \cite{Bonart} 
has shown that the default rate can diverge after a finite time, yet another 
signal of the
collective synchronization effects studied here. Second, one can consider an 
epidemic model \cite{Epidemic} where $x_i$ gauges the level of infection of individual $i$. 
When 
$x_i > \Th$, the illness declares itself and the disease becomes strongly 
contagious. The flux $J(t)$ would then model non-infected new immigrants in the 
population.
The model then predicts the possibility of sporadic outbreaks of the epidemics. 
Yet another interpretation of our model is in terms of the fiber bundle model 
for 
fracture \cite{Hansen}, allowing for ``self-healing'', i.e. the possibility for 
broken links to reform in time. Finally, one can interpret Eq. (\ref{eq:FP}) as a 
mean-field description of the 
depinning transition \cite{Depinning}, where each particle $i$ is subject to an 
increasing force $f_i=x_i$ until $f_i$ reaches a local depinning threshold 
$\Theta$. At this point, the
particle advances and thereby relaxes the force acting on it, and gets trapped 
again. However, if one assumes that particle $i$ is elastically coupled to its
neighbours $j$, the forces $f_j$ will increase as particle $i$ depins. In 
mean-field, we once again end up with Eq.(\ref{eq:FP}), which predicts a 
transition 
from a smooth overall progression when the external drive (here modeled by the 
drift $b$) is large enough to a jerky stick-slip motion for small drive. An 
interesting generalization is to consider that the feedback does not affect the 
drift $b$, as above, but the diffusion constant $D$, much as in \cite{Hebraud}. 
This could be relevant for soft
glassy matter or granular materials, where a localized yield event is often 
supposed to act as an effective temperature for the rest of the system. This, 
and 
other extensions, are left for future investigations. 

{\it Acknowledgements} We thank J. Bonart, E. Bouchaud and J. Donier for very insightful conversations.
This work was partially financed by the EU ``CRISIS'' project (grant number: FP7-ICT-2011-7-288501-CRISIS)


\begin{thebibliography}{99}
\bibitem{footnote} For model I, the final equation is even simpler and
reads: $\b \phi_0 = (1 - K D/b_0)/(1-e^{K \Th})$.
\end{thebibliography}
\end{document}